%% file: main.tex
\documentclass[11pt]{article}  
\usepackage{amsmath,amsfonts,amssymb}
\usepackage{graphicx}
\usepackage{setspace}
\usepackage{tocloft}
\usepackage{siunitx}
\usepackage{fullpage}
\usepackage{hyperref}

\hypersetup{
    colorlinks=true,
    linkcolor=blue,
    filecolor=magenta,      
    urlcolor=blue,
    pdftitle={Overleaf Example},
    pdfpagemode=FullScreen,
    }

\usepackage[margin=10pt,font=small,labelfont=bf]{caption}
\usepackage{xcolor}
\usepackage[noabbrev]{cleveref}

\input{macros}

\title{Enhancing signal detectability in learning-based CT reconstruction with a model observer inspired loss function}
\author{Megan Lantz, Emil Y. Sidky, Ingrid S. Reiser, Xiaochuan Pan, Gregory Ongie \footnote{M.~ Lantz and G.~Ongie are with the Department of Mathematical and Statistical Sciences, Marquette University, Milwaukee, WI, USA. E.Y.~Sidky, I.~Reiser, X.~Pan are with the Department of Radiology, University of Chicago, Chicago, Illinois, USA. Corresponding author: Gregory Ongie (\url{gregory.ongie@marquette.edu}) } }

\begin{document} 
\maketitle

\begin{abstract}
Deep neural networks used for reconstructing sparse-view CT data are typically trained by minimizing a pixel-wise mean-squared error or similar loss function over a set of training images. However, networks trained with such pixel-wise losses are prone to wipe out small, low-contrast features that are critical for screening and diagnosis. To remedy this issue, we introduce a novel training loss inspired by the model observer framework to enhance the detectability of weak signals in the reconstructions. We evaluate our approach on the reconstruction of synthetic sparse-view breast CT data, and demonstrate an improvement in signal detectability with the proposed loss.
\end{abstract}

\begin{spacing}{2}  

\section{Introduction}
\label{sect:intro}

\begin{figure}[ht!]
     \centering
     \includegraphics[width=0.7\columnwidth]{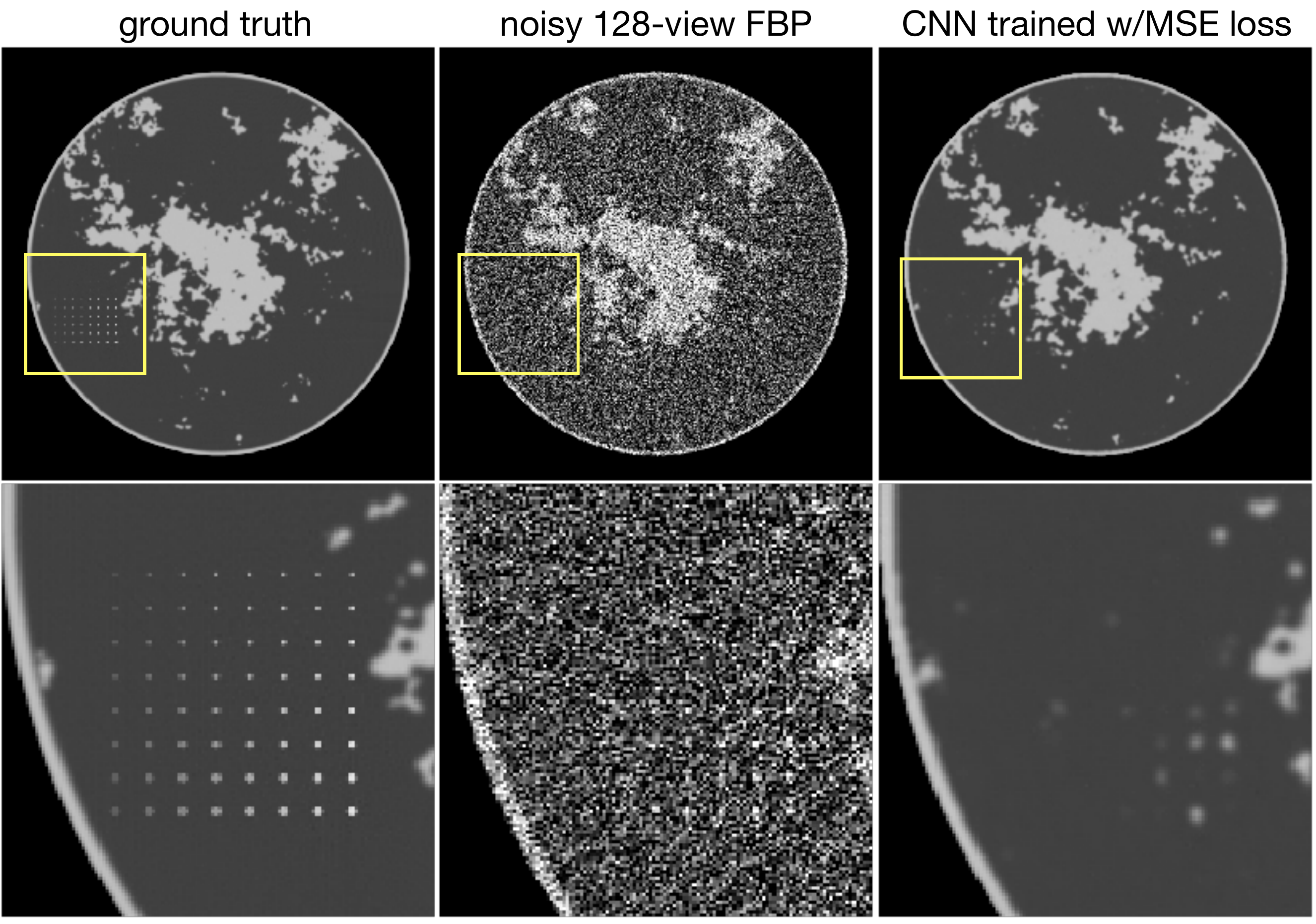}
     \caption{A CNN trained to perform sparse-view CT reconstruction by minimizing a pixel-wise mean-squared-error (MSE) loss is prone to wiping-out weak ``signals'' in a breast CT phantom. Left panel shows the ground truth phantom with a contrast-detail insert. Middle panel shows the simulated noisy, sparse-view FBP reconstruction. Right panel shows the reconstruction with a CNN trained with a pixel-wise MSE loss taking the noisy FBP image as input. Note many of the weak signals in the contrast-detail insert are clearly visible in the FBP image, but missing from the CNN reconstruction. More information on the data, network architectures, and training protocol used to generate this figure is given in \Cref{sec:methods}. All images are shown on the scale $[0.174, 0.253] \text{cm}^{-1}$.}
     \label{fig:pure_denoiser_demo}
 \end{figure}

Deep learning-based reconstructions of low-dose/sparse-view CT data \cite{jin2017deep,kang2017deep}, including breast CT \cite{cong2019deep}, have gained significant attention in recent years. Most current approaches train a convolutional neural network (CNN) to reconstruct CT images by minimizing a pixel-wise loss function, such as mean-squared error (MSE), over a set of training images. However, pixel-wise losses are insensitive to small and/or low-contrast features that may be critical for screening and diagnosis (e.g., microcalcifications and spiculations in the case of breast CT imaging) and these important features are likely to be lost in the reconstructions. See \Cref{fig:pure_denoiser_demo} for an illustration of this effect in the reconstruction of a simulated breast CT phantom. 

The source of this discrepancy is that pixel-wise losses are not inherently well-aligned with \emph{signal detection tasks} arising in medical imaging. Indeed, a variety of studies indicate that training CNNs to minimize a pixel-wise loss measure does not necessarily improve signal detection performance in image restoration or reconstruction\cite{kim2019performance,li2021assessing,yu2022investigating}. For example, these studies show that CNNs trained with the MSE loss can act as a low-pass filter, reducing the visibility of high-frequency elements in the reconstructed images and diminishing signal detectability performance \cite{yu2022investigating}. 

Recent research has explored alternatives to pixel-wise loss functions, including perceptual losses and adversarial losses. A perceptual loss uses intermediate-layer features of a pre-trained classification network (typically VGG16 \cite{simonyan2014very}) as a means to compare images \cite{johnson2016perceptual}. The name ``perceptual loss'' stems from the ability of this approach to identify semantic similarities between images. In a CT imaging context, a recent study showed that using a perceptual loss rather than a pixel-wise loss to train a CNN denoiser better preserved mid- and high-frequency features in denoised images, and showed improved detectability of fine signals \cite{kim2019performance}. Additionally, training a CNN denoiser using a Generative Adversarial Network (GAN) based adversarial loss in combination with a perceptual loss (as originally proposed by Yang et al. \cite{yang2018low}) has been shown to further improve signal detectability\cite{kim2019performance}.

The alternatives to pixel-wise loss functions discussed above demonstrate potential for enhancing signal detectability, but they do not explicitly focus on optimizing for such tasks. Recently, some studies have introduced specialized loss functions explicitly tailored to improve performance in signal detection tasks.
In 2021, Han \emph{et al.} proposed training a CNN denoiser using a loss function derived from a classification CNN trained on a binary signal detection task, and demonstrated the ability of this approach to preserve high-frequency components and improve signal detectability performance in reconstructed CT images \cite{han2021low,han2022perceptual}. Tivnan \emph{et al.} take a different approach to incorporating task-specific information into CNN training \cite{tivnan2023tunable}.  Starting from a generalized bias-variance decomposition of the MSE loss, they construct a loss function that incorporates properties of a desired signal to be detected by allowing parameterized manipulation of the bias and variance components of the loss function. When applied to binary signal detection tasks on images, this approach was demonstrated to outperform a CNN trained with a standard MSE loss.

Aligned with these recent efforts, we propose a novel loss function that is designed to promote the detectability of weak signals in reconstructed images.  Our proposed loss term is inspired by the model observer framework \cite{barrett2013foundations}. In particular, the loss term we propose approximates the signal detection performance of a model observer on a synthesized signal-known-exactly/background-known-statistically (SKE/BKS) task. We evaluate the proposed loss term on the reconstruction of synthetic sparse-view breast CT data. Our results indicate that signal detection performance improves with the proposed loss function, while only modestly impacting the MSE of the reconstructed images relative to standard training techniques.

Preliminary results of this study were presented in a conference paper \cite{ongie2022evaluation}.  This work extends our previous study by considering both fixed and random signal locations in training, calculating variance estimates of the signal detection metrics, and by investigating an alternate observer model to guide tuning parameter selection.

\section{Methods}
\label{sec:methods}
In this study, we focus on a supervised learning setting where a CT reconstruction map is learned by training an image-to-image CNN on matched pairs of noisy, sparse-view filtered back-projection (FBP) and noise-free ground truth images. We call the CNN being trained the \emph{reconstruction network}. After training, we test the ability of the reconstruction network to preserve weak signals by measuring the signal detection performance of model observer on a signal-known-exactly/background-known-exactly task using a signal model similar to the one used in training. Below we describe the data, network architectures, and proposed training and testing procedures in more detail.

\subsection{Data}
\label{sec:methods:data}
We focus on simulated breast CT data in this study. A stochastic breast model is used to
generate multiple phantoms for training and testing.
Random phantom images are generated using a structured fibroglandular tissue model \cite{reiser2010task,sidky2020cnns}. Adipose tissue and fibroglandular tissue are modelled as having an attenuation of $0.194 \, \mathrm{cm}^{-1}$ and $0.233 \, \mathrm{cm}^{-1}$, respectively. An initial image is generated on a $2048\times 2048$ pixel grid, representing an $18~\mathrm{cm}\times 18~\mathrm{cm}$ area in physical units. The phantom lies inside an inscribed $9 ~\mathrm{cm}$ circular field-of-view (FOV). Noisy sparse-view sinogram data is simulated under a 2D circular, fan-beam scanning geometry, which is representative of the mid-plane slice of a 3D circular
cone-beam scan. Specifically, fan beam projections along 128 equi-spaced views in a full 360 degree arc with 1024 detector pixels per view are simulated assuming total of $\num{4e10}$ incident photons, or approximately $n = 61,035$ photons incident per detector pixel. An approximation to Poisson noise is simulated by modeling the intensity at each sinogram pixel as Gaussian distributed with mean $s$ and variance $(n\exp(-s))^{-1}$, where $s$ is the ground truth sinogram intensity computed from the noise-free image. Here, we use Gaussian noise in place of Poisson noise in order to facilitate computation of the signal detection performance metrics.

\begin{figure}[ht!]
    \centering
    \includegraphics[width=0.95\columnwidth]{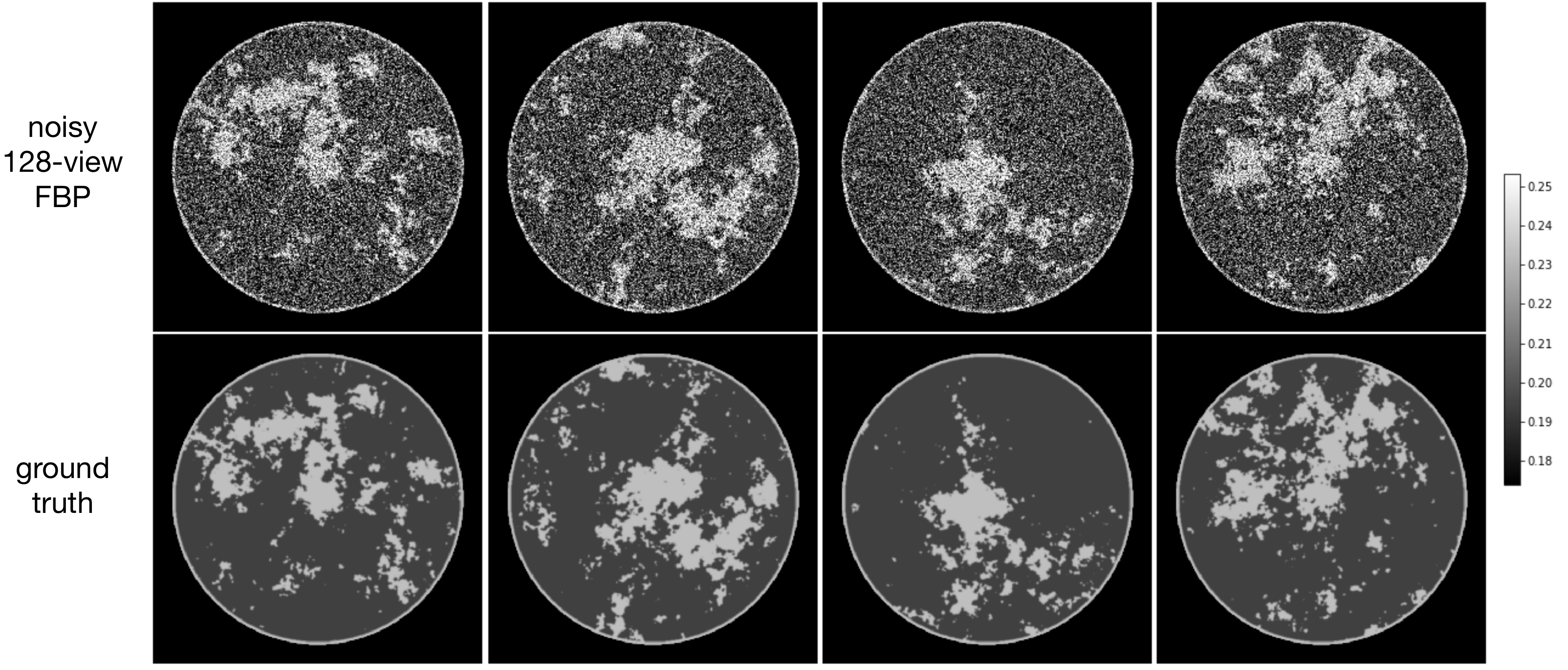}
    \caption{Examples of training pairs used in this study. Top row: simulated noisy 128-view FBP images. Bottom row: corresponding ground truth phantoms. All images are shown on the scale $[0.174, 0.253] \text{cm}^{-1}$.}
    \label{fig:train_data}
\end{figure}

Noise-free ground truth images are formed by downsampling the initial image to a $512\times 512$ pixel grid. Additionally, an initial FBP reconstruction on $512\times 512$ pixel grid is generated from the simulated noisy, sparse-view sinogram data, which is passed as input to the reconstruction network. A total of 1000 FBP and ground truth image pairs are used training; see Figure \ref{fig:train_data} for examples. Preparation of the test set is similar, but with some key differences described below.

\subsection{Network Architecture} 
In all our experiments, we use a U-net based CNN architecture \cite{ronneberger2015u} for our reconstruction network. Specifically, we use a U-net with 4 pooling layers, $3\times 3$ convolution kernels, and 32 channels per layer; dropout was not used. We modify the standard U-net architecture slightly by including an additional residual ``skip'' connection from the input layer to the output layer with trainable weights. More precisely, if $u_{\bthe'}(\cdot)$ denotes a standard U-net with trainable parameters $\bthe' \in \R^{p}$, we define our reconstruction network $f_{\bthe}(\cdot)$ as 
\begin{equation}\label{eq:alphabeta}
f_{\bthe}(\by) = \alpha \cdot \by + \beta\cdot u_{\bthe'}(\by),
\end{equation}
where $\by \in \R^d$ is a noisy, sparse-view FBP image, and $\bthe = (\alpha,\beta,\bthe')$ with $\alpha,\beta \in \mathbb{R}$ is the new set of trainable parameters. This formulation is similar to a residual learning strategy \cite{zhang2017beyond}, except we additionally allow the weight on the identity component to vary.

\subsection{Proposed loss function: the signal promoter}
Let $\{(\bx_n,\by_n)\}_{n=1}^N$ be a collection of training pairs, where each $\by_n \in \R^d$ is a noisy, sparse-view FBP image and $\bx_n \in \R^d$ is the corresponding ground truth image. 
We train the parameters $\bthe$ of the reconstruction network $f_{\bthe}$ by attempting to solve the following optimization problem:
\begin{equation}\label{eq:proposedloss}
\min_{\bthe} \frac{1}{N}\sum_{n=1}^N \|f_{\bthe}(\by_n)-\bx_n\|^2 + \lambda \cdot \text{SigPro}(f_{\bthe}),
\end{equation}
The first term is the pixel-wise MSE averaged over the training set and $\text{SigPro}(\cdot)$ is the \emph{signal promoter} loss function (defined in Eq.\,\eqref{eq:obsreg} below), which is designed to promote preservation of weak signals in the reconstructions. Here $\lambda > 0$ is a tunable parameter trading-off between the MSE loss and the signal promoter loss, such that larger values of $\lambda$ encourage promotion of weak signals in the reconstruction at the expense of the MSE loss.

The signal promoter loss is defined with respect to a user-specified collection of signals to be planted within the training images, which we call \emph{training signals}. In particular, we assume that $K$ input-output pairs $\{(\bs_k,\hat{\bs}_k)\}_{k=1}^K$ have been generated, where $\bs_k \in \R^d$ is the  signal in input space (e.g., its sparse-view FBP) and $\hat{\bs}_k \in \R^d$ is the same signal represented in output space (e.g., its gridded reconstruction). We then define $\text{SigPro}(f_{\bthe})$ for any reconstruction network $f_{\bthe}$ as
\begin{equation}\label{eq:obsreg}
\text{SigPro}(f_{\bthe}) = -\frac{1}{KN}\sum_{k=1}^K\sum_{n=1}^N\hat{\bs}_k^\top\big(f_{\bthe}(\by_n + \bs_k)-f_{\bthe}(\by_n)\big)
\end{equation}
The signal promoter loss measures the negative correlation between the true signal $\hat{\bs}_k$ and the difference between reconstructions with signal present and signal absent $f_{\bthe}(\by_n + \bs_k)-f_{\bthe}(\by_n)$, averaged over all training images and training signals. Minimizing this quantity maximizes the positive correlation, which intuitively should enhance signal detectability in the reconstructed images. 

Another way to view the signal promoter loss is as an approximation to the signal-to-noise ratio (SNR) of a linear test statistic that uses the signal image as its template in a signal-known-exactly/background-known-statistically (SKE/BKS) detection task \cite{barrett2013foundations}.  More precisely, for a fixed training signal pair $(\bs_k,\hat{\bs}_k)$, consider testing the hypothesis that the signal is absent ($H_{0,k}$) against the hypothesis that the signal is present in the reconstruction ($H_{1,k}$):
\begin{align*}
    H_{0,k}: \hat{\mathbf{x}} & = f_{\bthe}(\mathbf{y}) \\
    H_{1,k}: \hat{\mathbf{x}} & = f_{\bthe} (\mathbf{y}+\bs_k),
\end{align*}
where $\hat{\mathbf{x}}$ is the reconstructed image, $f_{\bthe}$ is the trained reconstruction network, $\mathbf{y}$ is the noisy sparse-view FBP reconstruction of a stochastic background image. Let $t:\R^d\rightarrow \R$ be any test statistic defined over reconstructions $\hat{\bx}$. The signal-to-noise ratio (SNR) of $t$ is defined as
\begin{equation}\label{eq:SNR}
\text{SNR}_t = 
\frac{\langle t \rangle_1 - \langle t \rangle_0}
{\sqrt{\frac{1}{2}\sigma_1^2 + \frac{1}{2}\sigma_0^2}}
\end{equation}
where $\langle t \rangle_i$ is the mean of $t(\hat{\bx})$ with signal present $(i=1)$ or with signal absent $(i=0)$, and $\sigma_i$ is the standard deviation of $t(\hat{\bx})$ with signal present $(i=1)$ or signal absent $(i=0)$. Taking $t_k$ to be the linear test statistic $t_k(\hat{\bx}) =\hat{\bs}_k^\top\hat{\bx}$, we have $\langle t_k \rangle_1 = \mathbb{E}[\hat{\bs}_k^\top f_{\bthe}(\by + \bs_k)]$ and $\langle t_k \rangle_0 = \mathbb{E}[ \hat{\bs}_k^\top f_{\bthe}(\by)]$, where the expectations are taken with respect to the joint distribution of noise and background. Therefore, by linearity of the expectation, we see that
\[
\langle t_k\rangle_1 - \langle t_k \rangle_0 = \mathbb{E}[\hat{\bs}_k^\top\big(f_{\bthe}(\by_n + \bs_k)-f_{\bthe}(\by_n)\big)].
\]
By replacing the expectation with the empirical mean over training inputs $\{\by_ n\}_{n=1}^N$, and averaging over all training signal pairs $\{(\bs_k,\hat{\bs}_k)\}_{k=1}^K$, we recover the signal promoter loss \eqref{eq:obsreg}.

Note that the signal promoter loss differs from the SNR formula \eqref{eq:SNR} in that it lacks normalization by the standard deviation terms. While in principle this normalization could be included in the definition \eqref{eq:obsreg}, we found that omitting it led to more stable training in our experiments, and so we focus on the un-normalized version in this work.

\subsection{Signal model}
\begin{figure}
    \centering
\includegraphics[width=0.7\columnwidth]{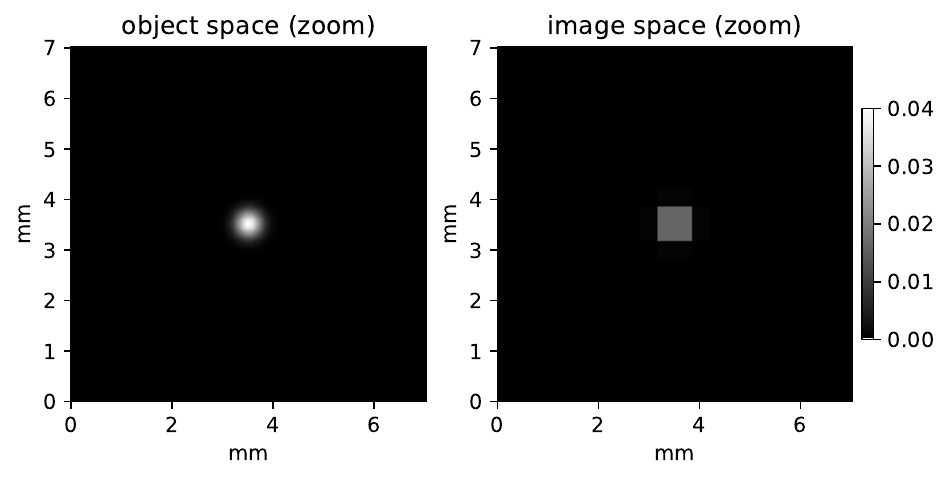}
    \caption{Illustration of signal model used in this study: (left) ideal Gaussian signal in object domain, (right) its counterpart in image space discretized to a $512\times 512$ pixel resolution. Both images are cropped to a $7~\textrm{mm} \times 7~\textrm{mm}$ region centered at the signal, which corresponds to a $20\times 20$ pixel region in image space.}
    \label{fig:signal}
\end{figure}

Rather than modelling a specific anatomical signal of interest, in our experiments we use ideal point-like training signals within a circular region-of-interest contained in the phantom. This is meant to act as a surrogate for a variety of signals relevant to detection tasks in CT applications. Specifically, we model each training signal in object domain as a Gaussian peak with an amplitude of 0.04 $\mathrm{cm}^{-1}$ and full width at half maximum of 1.25 pixels (0.44 mm). We use analytical expressions to generate a simulated sparse-view sinogram data of the signal, and use a Gaussian windowed version of an FBP reconstruction as the corresponding truth signal $\hat{\bs}$; see Figure \ref{fig:signal} for an illustration. 
Assuming the signal is located at the center of the FOV, and using a fixed background phantom, the area under the receiver operating curve (AUC) of the ideal data domain observer is roughly 0.98 on the corresponding signal-known-exactly/background-known-exactly (SKE/BKE) task. We consider two variations in the collection of training signals: $K=1000$ training signals centered at distinct random locations within the phantom anatomy, and $K=1$ training signal at a fixed single location near the center of the FOV. 

\subsection{Training procedure}
Incorporating the signal promoter loss \eqref{eq:obsreg} into standard gradient-based training of a reconstruction network is relatively straightforward. The signal promoter loss is linearly separable over the training images, which makes it compatible with stochastic gradient methods. In particular, when using training signals at random locations, we propose approximating the outer sum in \eqref{eq:obsreg} as a sum over mini-batches of training signals. For all of our experiments, we simply use a mini-batch size of one (i.e., add only one training signal per batch of training images).

Rather than optimizing \eqref{eq:proposedloss} from a random initialization of the weights of the reconstruction network, we find that a more effective strategy is to initialize with weights from a reconstruction network pre-trained with the MSE loss only (i.e., \eqref{eq:proposedloss} with $\lambda = 0$).  We then locally optimize the network weights to minimize the MSE loss combined with the signal promoter loss as in \eqref{eq:proposedloss}. In particular, we pre-trained a reconstruction network with paired noisy FBP and ground truth images to minimize the MSE loss using the Adam optimizer with a learning rate of $10^{-3}$ and batch size of 2 for 10 epochs. Then we fine-tune the reconstruction network by minimizing the loss in \eqref{eq:proposedloss} using a smaller learning rate of $10^{-5}$ and a batch size of 8 for one additional epoch. During the fine-tuning process, we freeze the weights of skip connection layer ($\alpha$ and $\beta$ in \eqref{eq:alphabeta}). Additionally, prior to the fine-tuning step, these weights are modified as $\alpha =  0.1 + 0.9\alpha_0$ and $\beta = 0.9\beta_0$, where $\alpha_0$ and $\beta_0$ are the pre-trained skip connection weights. This has the effect of adding back in $10\%$ of the input noisy FBP image to the CNN output. We observed that this modification helped the network weights deviate from their initialization during the fine-tuning stage, allowing the signal promoter loss to have a more immediate impact.

\begin{figure}
    \centering
    \includegraphics[width=\textwidth]{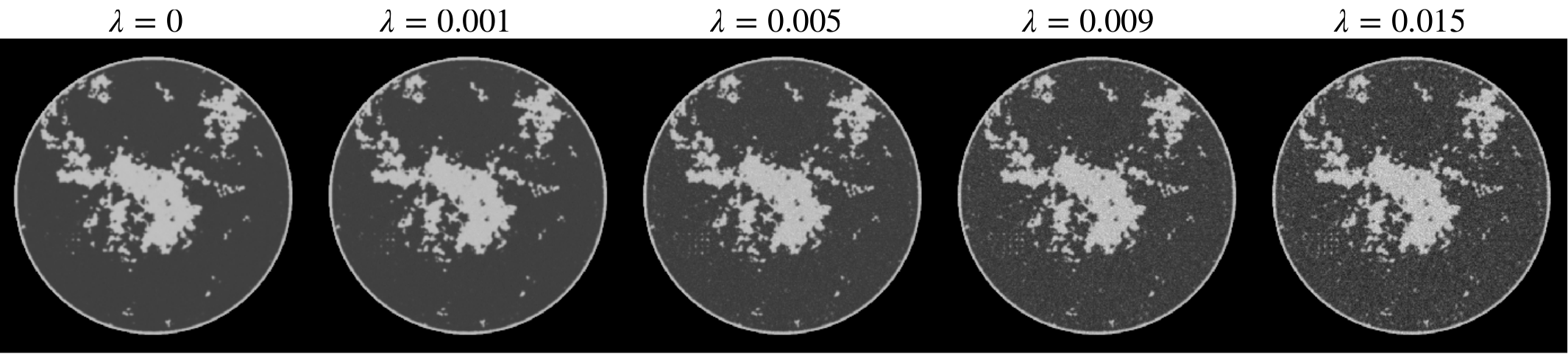}
    \caption{Reconstructions of a test phantom using CNN trained with the proposed signal promoter loss. Far left shows the reconstruction obtained from a CNN trained with MSE loss only, and with increasing weighting factor $\lambda$ on the signal promoter loss to the right. Note that more fine detail and noise becomes visible in the reconstructions as $\lambda$ increases. All images are shown on the scale $[0.174, 0.253] \text{cm}^{-1}$.}
    \label{fig:varylambda}
\end{figure}

We trained a series of 20 networks over a range of $\lambda$ values between $0$ and $0.20$ using the training procedure outlined above.  A $\lambda$ value near $0$ gives the signal promoter loss less weight and yields a reconstruction very similar to the pre-trained denoising network reconstruction, while higher values of $\lambda$ give more weight to the signal promoter loss and thereby put more emphasis on signal detection in the network training process. \Cref{fig:varylambda} illustrates this impact for a single test image reconstructed using five different reconstruction networks that span a range of $\lambda$ values.

\subsection{Testing procedure} 
 
For the testing phase, we set up a signal-known-exactly/background-known-exactly (SKE/BKE) task to measure signal detectability performance. This differs from the task used in training in that we consider a fixed background image and a slightly different signal model. Specifically, we test the hypothesis that the signal is absent ($H_0$) against the hypothesis that the signal is present in the image ($H_1$):
\begin{align*}
    H_0: \hat{\mathbf{x}} & = f_\theta (\mathbf{y_b+n}) \\
    H_1: \hat{\mathbf{x}} & = f_\theta (\mathbf{y_b+n+s}),
\end{align*}
where $f_{\bthe}$ is the trained reconstruction network, $\hat{\mathbf{x}}$ is the reconstructed image, $\mathbf{y_b}$ is the sparse-view FBP reconstruction of the noise-free sinogram of a fixed background image, $\mathbf{s}$ is the sparse-view FBP reconstruction of the noise-free signal only sinogram, and $\mathbf{n}$ is the sparse-view FBP reconstruction of Poisson-like additive Gaussian noise in sinogram domain.

\begin{figure}[ht!]
    \centering
    \includegraphics[width=0.7\columnwidth]{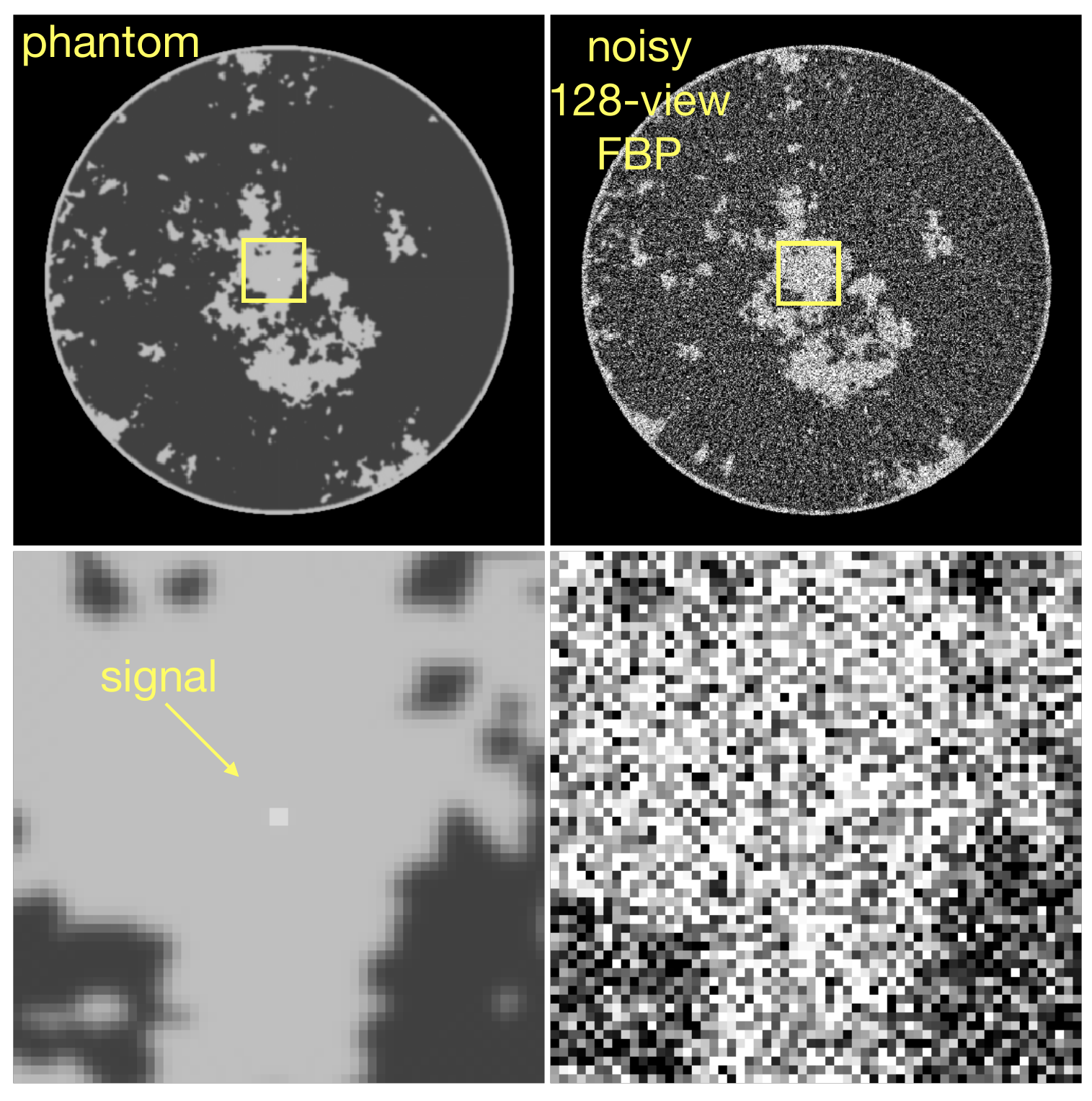}
    \caption{Illustration of signal-known-exactly/background-known-exactly task used in assessing signal detectability performance of reconstruction approaches. All images are shown on the scale $[0.174, 0.253] \text{cm}^{-1}$.}
    \label{fig:test_data}
\end{figure}

The test set consists of $1000$ signal present realizations and $1000$ signal absent realizations, all sharing the same fixed background image; see Figure \ref{fig:test_data} for an illustration. To facilitate computation of the signal detectability metrics, we fix the location of the test signal to the exact center of the image. We note that this location was not used for any of the training signals. The signal amplitude is set so that area under the receiver operating characteristic curve (AUC) of the ideal observer in sinogram domain is 0.86. This is half the amplitude of the training signals used in defining the signal promoter loss. 

As our figure-of-merit, we use the AUC as measured by a model observer in image domain. A classical AUC calculation is challenging in this setting because the reconstruction map is nonlinear.  Instead, we empirically estimate the AUC from reconstructed images using a channelized Hotelling observer (CHO) with a hybrid of pixel and Laguerre-Gauss channels, which have been found to be effective in measuring signal detection performance of nonlinear reconstruction methods \cite{medphy}. We use a total of 14 channels, with 4 individual pixel channels corresponding to a $2\times 2$ grid of pixels centered on signal, plus 10 radially symmetric Laguerre-Gauss channels. We use half of the test images to estimate the CHO coefficients, then we apply the estimated CHO to the remaining half of the reconstructed test set images to compute an empirical estimate of the AUC. We also report the MSE of the network over the entire test set.

To estimate the variance in the AUC of the CHO, we use a bootstrapping approach \cite{gallas2003variance}.  For each model, we simulate $100$ resampled datasets by randomly sampling, with replacement, from the test set images.  For each of the $100$ resampled datasets (each with $1000$ image pairs), the CHO is estimated and empirical estimates of the AUC and test set MSE are calculated as described above.  The mean and standard deviation of the test set MSE and AUC measures across the 100 re-sampled datasets is then calculated.

\section{Results}
As a baseline, we first train a reconstruction network using only the MSE as the loss function. Additionally, we consider an \emph{add-back-the-noise} strategy, where we form a convex combination of the initial noisy FBP image $\by$ with the output $\hat{\bx}_{0}$ of the reconstruction network trained with MSE loss only, i.e., we take the reconstructed image to be $\hat{\bx} = \alpha\by + (1-\alpha)\hat{\bx}_{0}$, where $\alpha \in [0,1]$ is treated as a tuning parameter. Note that ``noise'' here refers to the speckle artifacts due to noisy sinogram data and the streak artifacts from the sparse-view sampling. Finally, we compare the baseline add-back-the-noise approach with the set of networks trained using the proposed signal promoter loss, either with training signals at random locations or training signals at a fixed location.

\begin{figure}[!h]
\centering
\includegraphics[width=0.8\columnwidth] {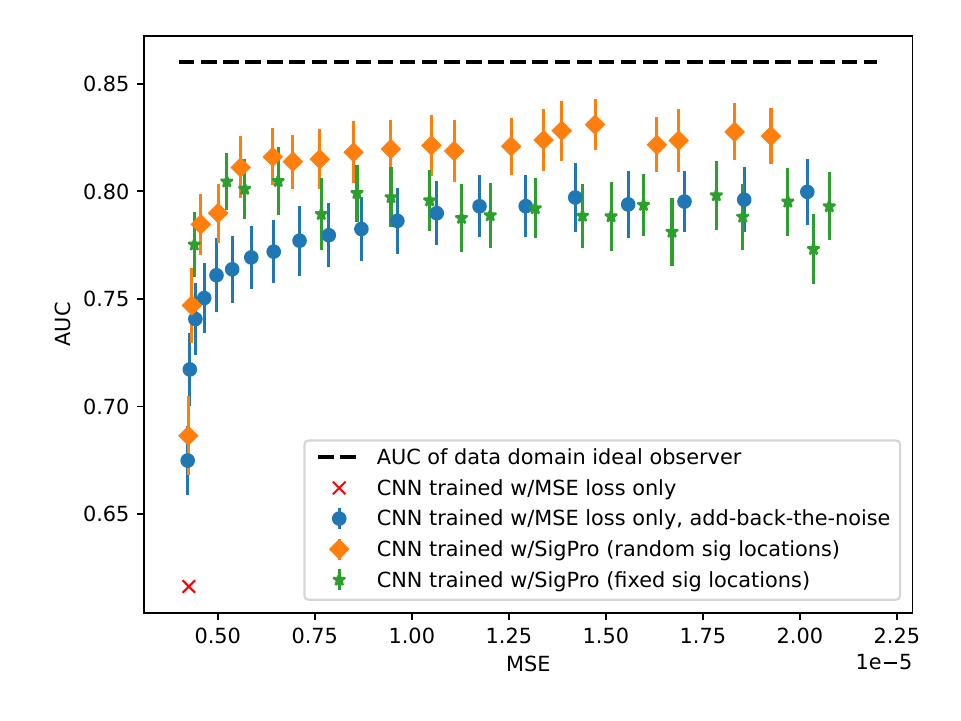}
\caption{Test set MSE and signal detectability performance (AUC) for various reconstruction networks. The MSE is measured over the full test set and the mean AUC of the channelized hotelling observer with respect to a SKE/BKE task is estimated empirically from reconstructed test images.  Vertical error bars indicate $\pm$ one standard deviation based on bootstrapped variance estimates.  
}
\label{fig:pareto}
\end{figure}
\subsection{Quantitative assessment of signal detection performance}
Our main results are shown in \Cref{fig:pareto}. For each reconstruction network we plot its MSE against AUC over the test set. Training with the MSE loss only gives the lowest MSE on the test set (MSE = \num{0.43e-5}), but also results in very low signal detectability (AUC = 0.53). This indicates that the corresponding CHO performs only slightly better than random guessing in detecting the signal. The add-back-the-noise strategy (in blue in Figure \ref{fig:pareto}) gives improvement in the AUC at the expense of moderate increases in MSE. In particular, the AUC monotonically increases with the fraction $\alpha$ of the initial noisy FBP image being added back in, saturating at an AUC of roughly $0.80$. The networks trained with the signal promoter loss show substantial improvements in signal detectability over this baseline. For a fixed MSE, the networks trained with the signal promoter loss using random training signal locations give uniformly higher AUC over the add-back-the-noise strategy.  When we consider small MSE's, the same is true for networks trained with the signal promoter loss using a single fixed training signal.  However, at higher levels of MSE, the add-back-the-noise baseline gives higher AUC than with the fixed training signal, indicating there is some loss of signal detectability for large signal promoter strengths $\lambda$. These results suggest that the signal promoter loss is effective at substantially increasing signal detectability performance while only moderately impacting MSE when the strength parameter $\lambda$ is small.

\begin{figure}[h!]
    \centering
    \includegraphics[width=0.95\columnwidth] {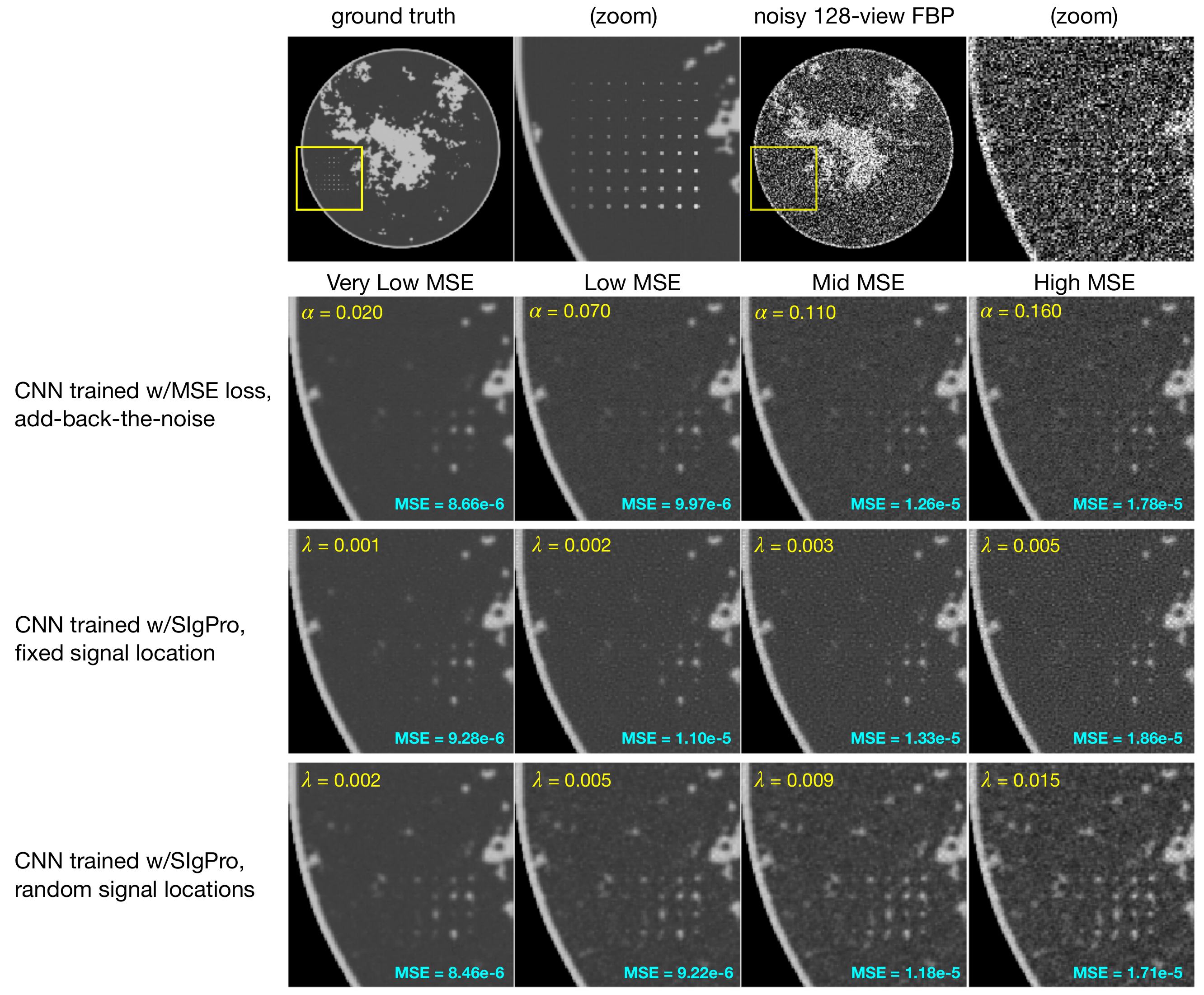} 
    \vspace{1em}
    \caption{Reconstructions of a test image that includes a contrast-detail insert. The first row shows the full and zoomed ground truth images and the noisy FBP input images for reference. The second row shows the output of a network trained to minimize MSE loss only with different fractions $\alpha$ of the noisy 128-view FBP image added back in. The third row shows the reconstructions obtained from networks trained with the signal promoter loss using a single training signal at a fixed location for various weighting factors $\lambda$.  Likewise, the fourth row shows the reconstructions obtained from networks trained with the signal promoter loss using random signal locations for various weighting factors $\lambda$. The reported MSE is the reconstruction error on the single displayed test image. Parameter values ($\alpha$ and $\lambda$) are selected such that images in each column have similar reconstruction MSE. All images are shown on the scale $[0.174, 0.253]\text{cm}^{-1}$. }
    \label{fig:mainfig}
\end{figure}

\subsection{Qualitative assessment using a contrast-detail insert}
In \Cref{fig:mainfig} we illustrate the correspondence between visual image quality and signal detectability metrics by reconstructing a single test image containing an additional contrast-detail (CD) insert. The CD insert consists of an $8\times 8$ grid of point-like signals of varying widths and contrasts. We compare reconstructions obtained by networks trained with the MSE loss only with the add-back-the-noise strategy, and networks trained with the signal promoter loss, both with random training signal locations and a fixed training signal location. For a fair comparison, we tune parameters so that a similar reconstruction MSE on the test phantom is achieved across all methods.  In \Cref{fig:mainfig}, the rows correspond to the reconstruction approaches and each column corresponds to a specific MSE budget.

Visually, we observe clear differences in the amount of the CD insert detail visible in reconstructions at the same MSE budget.
The reconstructions from networks trained with the signal promoter loss using both a fixed training signal and random training signal locations appear to preserve more signals in the contrast detail insert over the add-back-the-noise approach. We also observe visual differences in the noise characteristics of the reconstructed images between approaches.  Networks trained with the signal promoter loss with random training signal locations appear to preserve more signals in the contrast detail insert, but also show more pronounced noise and, at higher levels of the weighting parameter $\lambda$, show some evidence of introducing spurious features in the reconstructed image (see the bottom row of \Cref{fig:mainfig}). On the other hand, in reconstructions from networks trained with the signal promoter loss using a fixed signal location, the noise has a more uniform appearance, with fewer spurious high-contrast features (see the third row of \Cref{fig:mainfig}.) So, despite the fact that training with fixed signals generally underperformed training with random signals in terms of our AUC metric, the visual comparisons suggest that using multiple training signals at random locations may introduce more false positives (i.e., spurious high-contrast features), than using a single training signal at a fixed location. 

\subsection{Parameter Selection}
When training with the signal promoter loss we generated several different reconstruction networks based on different values of the weighting factor $\lambda$. In general, as $\lambda$ increases, we found that both the test set MSE and signal detectability performance (as measured by the AUC of the CHO) tend to increase. This phenomenon suggests that there is no optimal trade off between MSE and signal detectability performance when using the signal promoter loss. That is, from the point of view of signal detection, the weight parameter $\lambda$ that should be as large as possible.  However, visual comparison of reconstructed images in \Cref{fig:mainfig} shows that overall image quality degrades beyond any acceptable clinical standard when taking $\lambda$ very large. This leaves open the question of how to ``optimally'' tune the weighting factor $\lambda$.

To illustrate one potential approach to selecting the weighting factor $\lambda$,  we applied an alternative model observer that is ``weaker'' than the CHO. In particular, we use a signal-Laplacian model observer proposed by Ongie \emph{et al.} \cite{ongie2022evaluation} as a potential proxy for a human observer. The signal-Laplacian observer uses a linear test statistic whose template is the discrete Laplacian of the matched signal. Since the discrete Laplacian acts as a band-pass filter, the signal-Laplacian observer is similar to applying a non-prewhitening eye filter to a template given by the matched signal \cite{burgess1994statistically}. 

\begin{figure}[h!]
    \centering
    \includegraphics[width=\columnwidth]{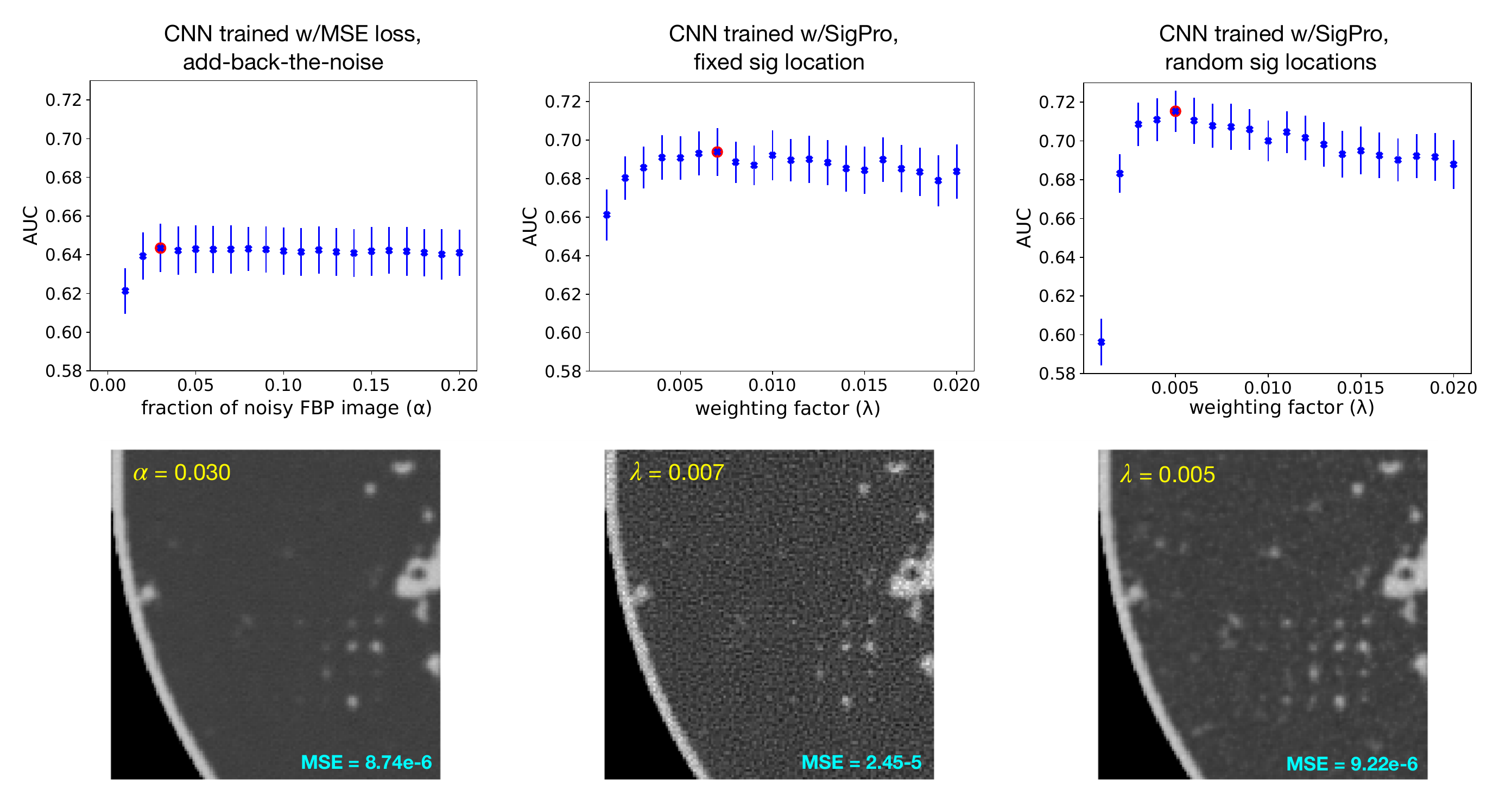}
    \caption{Top row: Tuning parameter ($\alpha$ or $\lambda$) vs. test set AUC of the signal-Laplacian model observer on a SKE/BKE detection task.  In contrast to the CHO AUC curves in \Cref{fig:pareto}, the signal-Laplacian model observer AUC appear to have a peak value (highlighted in red).  
    Bottom row: Reconstructions of the test image with an added CD insert with the ideal weighting parameter based on the maximum AUC of the signal-Laplacian observer. All images are shown on the scale $[0.174, 0.253] \text{cm}^{-1}$.}
    \label{fig:param_select}
\end{figure}

In \Cref{fig:param_select} we show the results of repeating the testing procedure described above using the signal-Laplacian observer in place of a CHO. We see that the AUC of the signal-Laplacian observer quickly peaks and then slowly tapers off as the weighting parameter $\lambda$ increases. We also show the same is true in terms of the add-back-the-noise approach with respect to its tuning parameter $\alpha$ (i.e., the fraction of noisy sparse-view FBP image added back in). The existence of this peak AUC value offers one potential method for selecting the ideal weighting parameter to use in training/selecting a reconstruction model.  

Based on the signal-Laplacian model observer, the ideal value of $\alpha$ for the add-back-the-noise approach is $\alpha = 0.030$, the ideal value of tuning parameter $\lambda$ for a network trained with the signal promoter loss and a fixed training signal is $\lambda = 0.007$, and $\lambda = 0.005$ when using random training signal locations. In \Cref{fig:param_select} we show reconstructions of the test phantom with the CD insert using the different models at these selected parameter values. 

\section{Discussion and Conclusion}
Our findings confirm that training CNNs to perform sparse-view CT reconstruction using a standard MSE loss function may lead to poor signal detectability performance, as measured by the performance of a model observer on a SKE/BKE detection task. We demonstrate that re-optimizing the CNN using a newly proposed signal promoter loss term in addition to the MSE loss leads to improved signal detectability with only modest increases in test set MSE. Finally, we demonstrate one approach for the selection of tuning parameters in the newly proposed loss function by assessing signal detection performance using a linear model observer based on a signal-Laplacian template.

Concurrent with this work, Li \emph{et al.}~recently proposed a task-based loss function inspired by the model observer framework for CT image reconstruction \cite{li2022impact}.  Li \emph{et al.}~use two variations on a model observer term in their proposed task-based loss function. Both of these approaches involve estimating the detection performance of a model observer on a SKE/BKE task during training. One approach uses the classification accuracy of (an approximation to) the ideal observer (IO) as a loss term. The IO is approximated using a CNN, which is trained in tandem with the reconstruction network using the binary cross-entropy loss. The other approach uses the signal-to-noise ratio (SNR) of the Hotelling observer (HO) as a loss term. The HO is approximated using a single-layer linear neural network, whose parameters are estimated in tandem with the reconstruction network using a variational characterization of the SNR of the HO \cite{zhou2019approximating}. Similar to our approach, Li \emph{et al.} start from a pre-trained reconstruction network and then refine the reconstruction network parameters using the model observer inspired losses.
However, our proposed approach differs from Li \emph{et al.} in that we use a fixed model observer during training, rather than attempting to estimate HO or IO during training. Additionally,  Li \emph{et al.} focuses on detecting tumors or nodules in low dose lung CT images, which is qualitatively different from the point-like signals investigated in this study.

While this work demonstrates improvements in signal detectability using the proposed signal promoter loss, there are limits to the scope of conclusions that can be drawn here. This study included only simulated data, but in principle, the signal promoter loss proposed here could be used when training a reconstruction network on real breast CT data, since it does not rely on specific properties of the training set. Also, we focused on a supervised learning setting in our experiments where access to paired noisy sparse-view FBP and ground truth images is assumed. However, since signal promoter loss does not require access to ground truth images, it can potentially be used in unsupervised or semi-supervised settings \cite{wang2020deep}, as well. 

Additionally, in our experiments, we focused on reconstruction networks having a feed-forward U-net CNN architecture \cite{ronneberger2015u}. However, the signal promoter loss function is agnostic to architecture, and can be applied to other state-of-the-art trainable estimators, such as those based on algorithm unrolling that iteratively enforce data consistency constraints \cite{monga2021algorithm}. Such unrolled estimators have recently shown promise in enhancing signal detectability in digital breast tomosynthesis reconstruction \cite{gao2023model}, and incorporating the signal promoter loss into training unrolled estimators may further enhance this effect.  

Finally, our experiments only considered training with the signal promoter loss in conjunction with the MSE loss, but other commonly-used losses could have been used in place of the MSE loss, such as the mean absolute error or structural similarity index measure \cite{wang2004image}, or even perceptual/adversarial losses \cite{han2022perceptual}. Exploring possible synergistic effects of combining the signal promoter loss with these other loss functions is an interesting direction for future work.

\subsection*{Disclosures}
The authors have no relevant financial interests or conflicts of interest to disclose.

\subsection* {Acknowledgments}
This work is supported in part by NIH Grant Nos. R01-EB023968 and R21-CA263660. The contents of this article are solely the responsibility of the authors and do not necessarily represent the official views of the National Institutes of Health.
Additionally, G.O. and M.L. were supported by NSF CRII award CCF-2153371.

\subsection* {Code, Data, and Materials Availability} 
Code to reproduce all results and figures in this paper is publicly available in the GitHub repository: \url{https://github.com/m-lantz/ct_recon_SigPro}


\bibliography{main} 
\bibliographystyle{ieeetr}

\end{spacing}
\end{document}

%% file: macros.tex
\newcommand{\bx}{\mathbf{x}}
\newcommand{\by}{\mathbf{y}}
\newcommand{\bs}{\mathbf{s}}

\newcommand{\R}{\mathbb{R}}

\newcommand{\bthe}{\boldsymbol{\theta}}